# Telerobotic Pointing Gestures Shape Human Spatial Cognition

John-John Cabibihan,[1,*] Wing-Chee So,[2] Sujin Saj[1], Zhengchen Zhang[1]

[1] *Social Robotics Laboratory, Interactive and Digital Media Institute; and the Department of Electrical and Computer Engineering, National University of Singapore*

[2] *Department of Educational Psychology, The Chinese University of Hong Kong*

[*]Corresponding author's contact: John-John Cabibihan (elecjj@nus.edu.sg)

Supplementary videos of the experiments can be found on the link below:
http://www.youtube.com/watch?v=HBVEmnfzX4c

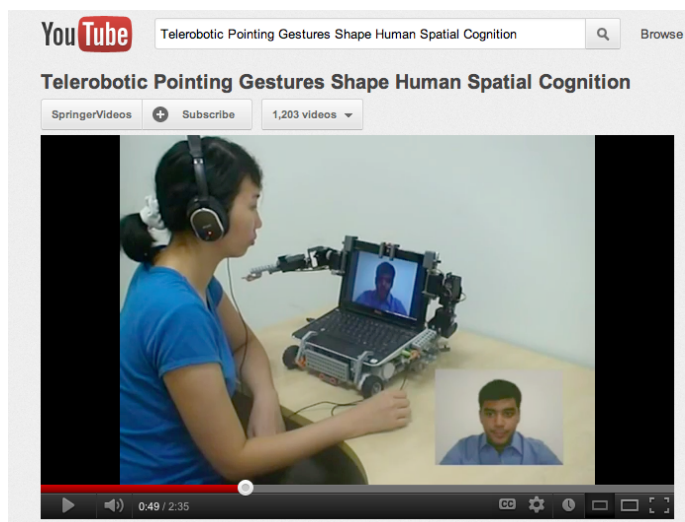

---

The final publication is available at springerlink.com

# Abstract

This paper aimed to explore whether human beings can understand gestures produced by telepresence robots. If it were the case, they can derive meaning conveyed in telerobotic gestures when processing spatial information. We conducted two experiments over Skype in the present study. Participants were presented with a robotic interface that had arms, which were teleoperated by an experimenter. The robot could point to virtual locations that represented certain entities. In Experiment 1, the experimenter described spatial locations of fictitious objects sequentially in two conditions: speech condition (SO, verbal descriptions clearly indicated the spatial layout) and speech and gesture condition (SR, verbal descriptions were ambiguous but accompanied by robotic pointing gestures). Participants were then asked to recall the objects' spatial locations. We found that the number of spatial locations recalled in the SR condition was on par with that in the SO condition, suggesting that telerobotic pointing gestures compensated ambiguous speech during the process of spatial information. In Experiment 2, the experimenter described spatial locations non-sequentially in the SR and SO conditions. Surprisingly, the number of spatial locations recalled in the SR condition was even higher than that in the SO condition, suggesting that telerobotic pointing gestures were more powerful than speech in conveying spatial information when information was presented in an unpredictable order. The findings provide evidence that human beings are able to comprehend telerobotic gestures, and importantly, integrate these gestures with co-occurring speech. This work promotes engaging remote collaboration among humans through a robot intermediary.

# Keywords

# Introduction

Telepresence refers to instruments that will allow us to work remotely in another room, in another city, in another country, or in another planet, and using these instruments will feel and work so much like our own hands that we would not notice any significant difference [1, 2]. Telerobotic pointing gestures, unlike verbal descriptions, locate and anchor entities in *space* [3]. They can simulate an image of spatial locations of entities in the listeners' mental representation [4]. In addition, it allows listeners to have a more elaborate encoding of the materials, resulting in a rich memory base for retrieval of the spatial locations [5].

Researchers have built certain telerobots in the past. For example, the telepresence robot, QB [6], was deployed in an office environment to investigate the benefits of a robot avatar as a stand-in for an office employee (Fig. 1A). The employee controls the robot from a computer at his home. One of QB's robotic eyes captures high-definition video while the other was designed to point a laser beam at things. The laser serves to compensate for QB's lack of robotic arms for pointing. Embodied as a robot, the remotely located office worker was able interact with colleagues as per normal. Other types of telepresence robots for human environments are emerging in the areas of medical consultations [7-12], long distance education [13-15] and social networking [16-18]. Moreover, telepresence robots are being used as research platforms to understand human-robot personal relationships [19-21].

Many telepresence robots that have been developed do not have pointing hands. Take Rhino for example (Fig 1B). This robot is an interactive museum guide robot that was deployed in a densely populated museum [22]. It was equipped with verbal, graphical and audio capabilities and it was responsible for giving visitors directions in a museum [23]. Because Rhino did not have pointing hands, it had to lead the way for the visitors to follow. In fact, Rhino could have easily pointed the directions similar to what humans typically do.

Another example is the telepresence robot by InTouch Health (Santa Barbara, CA). Telementoring has opened the possibility for expert surgeons to assist another surgeon from a remote location due to an increase of travel expenses [24-27]. Recently, telepresence robots have been used to act as an intermediary between the mentor and the mentee [28-30]. The robots are life-sized, wheeled robots that can be controlled by a surgeon to maneuver in a hospital floor. The person in front of the robot can interact with surgeon. The robot's head is a flat monitor that can rotate (Fig. 1C). However, it does not have arms. As with other forms of transferring a skill, demonstrations using verbal instructions together with hand motions and pointing gestures are critical for learning. For remote instructions through the web, pointing gestures can provide additional help because it reduces the reference frame conflicts (cf. [31-34]). Specifically, the spatial locations of the objects can be described in relation to the location of the listener or to another object.

Pointing is a fundamental building block of human communication [35]. The presence of pointing gestures is particularly essential for listeners when co-occurring speech is ambiguous [36], e.g., a speaker says, "The door is on that side." It is unclear for listeners where the door is. However, a speaker can use a pointing gesture (e.g., points to his left) to disambiguate speech. To investigate the effects of pointing gestures on the human observer, we designed a telepresence robot with pointing hands and instructed it point to spatial locations that are associated with entities. We asked whether human beings can comprehend telerobotic pointing gestures. If so, encoding these pointing gestures should enhance humans' spatial representation, particularly when co-occurring speech is ambiguous.

In the present study, we conducted two experiments over Skype (Skype Ltd, Luxembourg) to address this question. In both experiments, the participants sat in front of a robotic interface. This robot has arms that can be teleoperated by the experimenter. The experimenter described the spatial locations of five fictitious objects in four typical rooms. The participants were then asked to recall the spatial locations of the objects by dragging them to the correct locations on a separate software. In Experiment 1, the objects' spatial locations in the room were described *sequentially* from left to right (Fig. 2A and 2B and Movies S1 and S2). In Experiment 2, the objects' spatial locations in the room were described in a non-sequential order (Fig. 2C and 2D and Movies S3 and S4). In each experiment, the participants received two modes of descriptions from the experimenter: speech only (SO condition) and speech with robotic pointing gestures (SR condition). In the SO condition, speech descriptions clearly indicated the spatial locations of objects, e.g., "At your left corner is the computer," and the robot was not instructed to gesture. In the SR condition, speech descriptions were ambiguous, e.g., "You can find the computer over there," but the robot was instructed to point to the participants' left. The order of the conditions was counterbalanced.

# Materials and Methods

### Participants

Forty subjects (32 males) participated in Experiment 1 and another group forty-two subjects (26 males) participated in Experiment 2. All participants were undergraduate or graduate students (18-30 years old) at the National University of Singapore. All were English speakers and have prior experience in using Skype. They were recruited by email and were reimbursed $8 each for their participation.

### Experimental Procedure

We selected five unique objects for each of the four rooms, i.e. kid's playroom, kitchen, living room and the bedroom. The results from the first room (i.e. kid's playroom) were discarded in the analysis due to the practice effects. To control for any effects arising from the experimenter's reading speed and tone of voice, we took the video of the experimenter while he was reading each text. The video for each room was 20 seconds in length. The participants were oriented on the experimental objectives and procedures as soon as they entered the experiment room. They sat in front of the telerobot interface and were asked to wear a set of noise cancelling headphones (SHN9500, Philips, Netherlands) to avoid any disturbance from the external environment or from the robot's mechanical movements. After the video for each room was viewed, the participants then moved to another computer with the custom-built software (Fig. 3) to position the objects according to the best of their memory. Each participant took about 12 to 16 seconds to complete this task. They returned in front of the robot interface to resume the viewing of instructions for the next room. A repeated measures ANOVA was used to analyze the data, which were run using Statistica (v10,

StatSoft, OK, USA).

**The Telerobotic Pointing Interface**

The robot was embodied with two arms having 3 degrees of freedom each (one at the elbow and two at the shoulder). The arms were attached on a netbook computer, which was then mounted above a mobile robot base. The robotic arms were assembled using 6 servo motors (HS-422, HiTec, CA, USA). These were controlled by a microcontroller (Mini Robotics MRK-4, A-Main Objectives, Singapore), which can sufficiently manage the 6 servo motors in a cascaded manner. The RS232 cable input of the microcontroller enabled the data communication between the microcontroller and the onboard netbook computer (Dell Inspiron Mini 10, 1.66GHz processor, 1 GB of RAM and a built-in web camera). The experimenter can remotely control the robot arms over the internet using custom-built software applications.

For external communications with the telepresence robot, a Structured Query Language server (SQL; MySQL, Community Server v5.5, Oracle) was setup on a dedicated server to host a shared database (Fig. 4). The SQL database serves as the main link between the experimenter's computer and the robot. This SQL database needed to be hosted on a dedicated server with a static IP address to ensure that the robot could function anywhere as long as it had internet access without having to consider factors such as being on the same network domain/subnet. The experimenter sends a pointing direction command to the SQL database. The software on the robot polls the database for new commands every 250 msec. If new commands were found, the software issues the command for the robot arm controller to execute. The software then updates the database to indicate that the command has been carried out.

The experimenter and the participant communicated via Skype. In a typical Skype videoconference, it is common that both parties would make use of their webcams and microphones to interact with each other. We recorded the video of the experimenter. We created the effect of real-time interaction by using Skype's Screen Share feature to display the experimenter's computer screen onto the participant's computer screen. We prepared the playlist of videos on the experimenter's screen and relayed this to the participant's screen.

**Texts**

In Experiment 1, the objects' locations in the room were described from left to right in a sequential order. The direction was consistently sweeping from one object to the next at an angular difference of 30° (Fig. 5A). The sum of the relative angular differences is 120° (Fig. 5B). The texts are given in the Appendix, left column. For the SR condition, the descriptions in parentheses refer to the directions of pointing gestures. In Experiment 2, the objects' spatial locations in the room were described in a non-sequential order. The spatial locations were arranged in a pseudo-random manner to minimize the effect of presentation order. For the SO condition for the children's playroom, kitchen, living room and the bedroom, the pseudo-random order was 51243, 24135, 13524 and 42351, respectively. Each digit corresponds to a location relative to the listener's egocentric frame of reference (i.e. 1 = participant's left corner; 2 = left side; 3 = behind the participant; 4 = right side; and 5 = right corner). For the SR condition, the order was 14235, 35421, 51342 and 23514 for those rooms. These correspond to angular differences of 210°, 180°, 270° and 270° (Fig. 5C to Fig. 5F). The texts for the non-sequential presentation of objects are shown in Appendix, right column. Likewise, the descriptions in parentheses refer to the directions of pointing gestures.

# Results

### Recall of Objects and their Spatial Locations

Experiment 1 (sequential) investigated whether the participants were able to comprehend the telerobotic pointing gestures. If so, they would be able to recall a comparable number of spatial locations in the SO and SR conditions. Otherwise, they would recall significantly fewer spatial locations in the SR condition than in the SO condition, since speech descriptions in the SR condition were ambiguous. Fig. 6 (left panel) shows the proportion of objects that were correctly recalled in the SO and SR conditions. There was no significant difference between the conditions, $[F(1,38) = 0.07, P = 0.79]$. In other words, the recall rate in the SR condition was on par with that in the SO condition. Hence, the findings supported the first hypothesis that participants comprehended the telerobotic pointing gestures. More importantly, they made use of the telerobotic pointing gestures to disambiguate co-occurring speech. There were no main effect for the order of condition $[F(1,38) = 0.75, P = 0.39]$ and interactions.

Experiment 2 (non-sequential) investigated whether the findings in Experiment 1 could be replicated. Fig. 6 (right panel) shows the proportion of objects correctly recalled in the SO and SR conditions. Interestingly, the participants recalled even more spatial locations in the SR condition than in the SO condition $[F(1,40) = 7.64, P = 0.0085]$. In other words, pointing gestures produced along with ambiguous speech even yielded better recall than unambiguous speech. This finding highlights that speech with telerobotic pointing gestures is more powerful than speech description alone in conveying spatial information and enhancing spatial recall when the sequence of descriptions of objects was unpredictable. The order of condition was not significant [$F(1,40)$ = 0.06, $P$ = 0.81] and there were no significant interaction effects.

**Effect of the Number of Words**

We verified whether the number of words played a role in the results shown in Fig. 6. We controlled the number of words in the text that the experimenter will read. The number of words that were used is as follows: kitchen, 38 words; living room, 39 words; and bedroom, 38 words. The proportion of correctly remembered objects for the three rooms is shown in Fig. 7. Each room corresponds to the number of words used in the text. For Experiment 1, the left panel shows no significant difference among the rooms [$F(2,76)$ = 0.13, $P$ = 0.88]. Similarly for Experiment 2, we did not observe any significant difference in the effect of the number of words in the text [$F(2,80)$ = 1.53, $P$ = 0.22]. The objects that were selected for each room were familiar objects. Taken together with the manipulation of the words in the texts and the pseudo-randomized sequence, these results rule out the possibility that the number of words was a confounding factor in the recall of objects and their locations.

# Discussion and Conclusions

The present study investigated whether the participants comprehended the telerobotic pointing gestures and whether encoding these pointing gestures strengthened their spatial memory. Our findings showed that these were indeed the case. Specifically, when co-occurring speech was ambiguous, the participants could derive spatial information from telerobotic pointing gestures and use it for spatial processing. Thus, the recall rate in the SR condition was similar to that in the SO condition. Such finding has two important implications. First, like human pointing gestures, telerobotic pointing gestures convey spatial information, and second, they are integrated with speech. More importantly, telerobotic gestures can disambiguate co-occurring speech when necessary.

Speech-associated gestures convey information that is meaningful to listeners [37, 38]. Henceforth, incorporating gestures in human-robot communication can evoke meaningful social interaction. Research on the interpretation of robotic hand gesture is relatively scarce but these earlier findings were consistent. Kanda et al. [39] found that human beings responded to body movements and utterances by a route guidance robot while Oberman et al. [40] further reported that comprehending robotic gestures might activate the mirror neuron system that was previously thought to be specifically selective for biological actions.

In this study, we demonstrated the role of telerobotic gestures in shaping human cognition, especially spatial cognition. We focused on pointing gestures in which the robot pointed to a spatial location that was associated with a particular entity. Our results were consistent with the previous findings in human-to-human interactions that showed that pointing gesture serves the function of conveying spatial information to the listeners [41-44]. Earlier studies showed that pointing simulates an image of the spatial locations of entities in the listeners' mental representation [4, 41, 45]. In addition, pointing allows the listeners to establish an alternative memory trace of spatial information in a nonverbal modality, together with the verbal modality. According to the Dual-Coding Theory [46, 47] and the Conjoint Retention Hypothesis [48-50], verbal and spatial information are stored separately. Therefore, the presentation of both types of information allows listeners to have a more elaborate encoding of the material, and thus, a rich memory base for retrieval [5].

Indeed, our findings suggest that pointing gestures produced by telepresence robots could facilitate language and cognitive processing in humans. Thus, pointing hands should become part of the design of telepresence robots in the future. A striking aspect of this study is that the participants were able to achieve higher recall scores when pointing gestures accompanied the verbal instructions—even when the spatial locations were presented in a non-sequential order. This finding may have practical applications for remote collaboration tasks in which critical information about objects and their spatial locations are communicated. The non-sequential order could be analogous to the way information is presented during emergency situations in unknown and hazardous environments. There could be minimal time to organize the information; hence, the sequence of the spatial locations that is being communicated will be unpredictable. Furthermore, it is conceivable that the behaviors for indicating objects or demonstrating procedures (cf. [51]) during face-to-face discourse will continue in web-based communications wherein pointing gestures will still be performed. Observational studies in video-based remote collaborations have earlier suggested that the two-dimensional field of view, which is the current standard in peer-to-peer video communications, is often not sufficient to show the full range of gestural motions in space; consequently, the "common ground" is not established [52-55]. Conflicting reference frames between the speaker and the listener can also arise and parts of the speaker's face can be blocked if explicit pointing motion is made. When programmed on the arm of the personal robot companions for the elderly or for children, the robot's pointing gestures can provide a quick and efficient way to communicate object-spatial information to the listener.

To conclude, our findings in both experiments show that participants were able to understand telerobotic pointing gestures and integrate them in co-occurring speech. More importantly, they also flexibly turned to the pointing gestures when they were exposed to a circumstance in which they could not predict what would come next. This work promotes a more engaging remote collaboration among humans through a robot intermediary.

# Acknowledgments

This work was supported by the National University of Singapore Academic Research Funding Grant No. R-263-000-576-112. Special thanks to Lwin Htay Thet and Nagasubramaniam Kumarappan for their assistance in the experiments and pilot studies.

# Figures

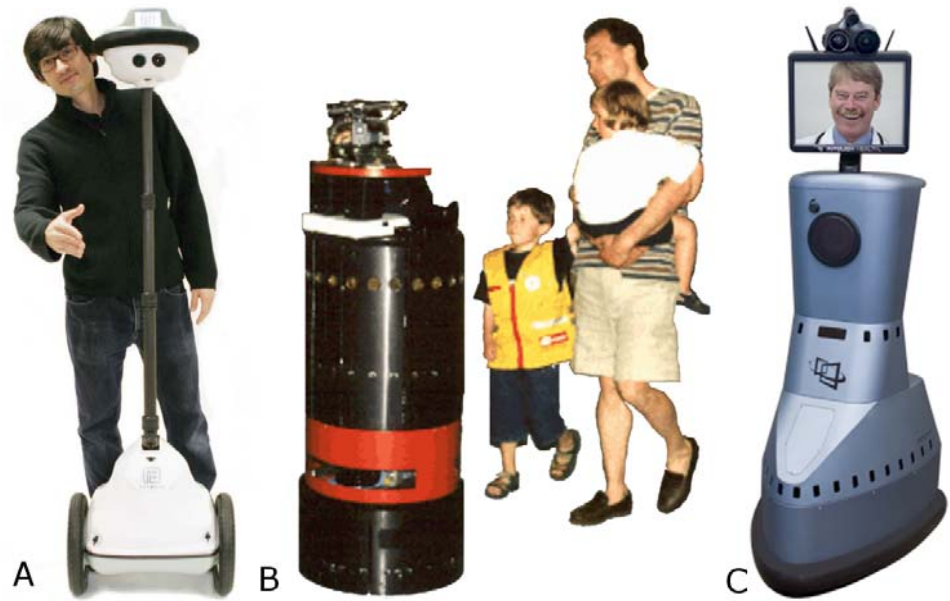

Fig. 1. Arm-less telepresence robots. (A) Anybot's QB personal avatar ©2010 IEEE. Reprinted, with permission, from E. Guizzo, "When my avatar went to work," IEEE Spectrum 47; (B) Rhino Museum Guide, image courtesy of Prof. P.E. Trahanias, "TOURBOT – Interactive Museum Telepresence Through Robotics Avatars," Cultivate Interactive, issue 2, Oct 2000; and (C) RP-7 robot, image courtesy of InTouch Health.

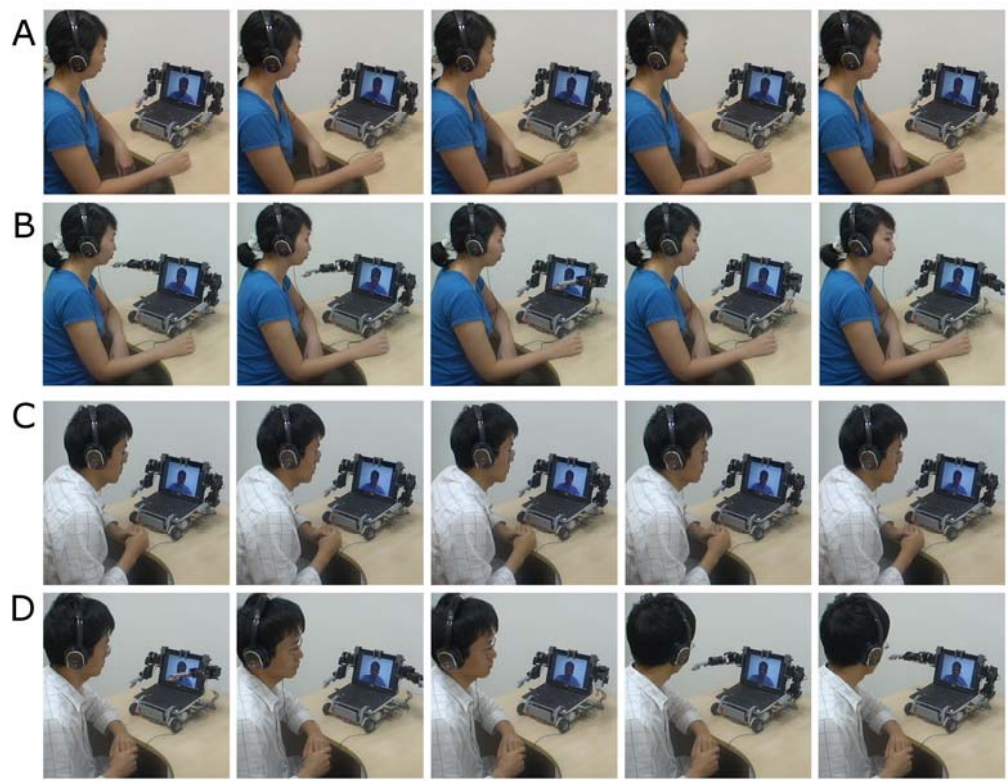

Fig. 2. Examples of the interactions between the participants and the experimenter through the telepresence robot. (Top two rows) The fictitious objects are presented in a sequential manner (A) using verbal instructions only and (B) using verbal and pointing gestures. (Bottom two rows) Non-sequential order of presenting the objects, (C) using verbal instructions only and (D) using verbal and pointing gestures.

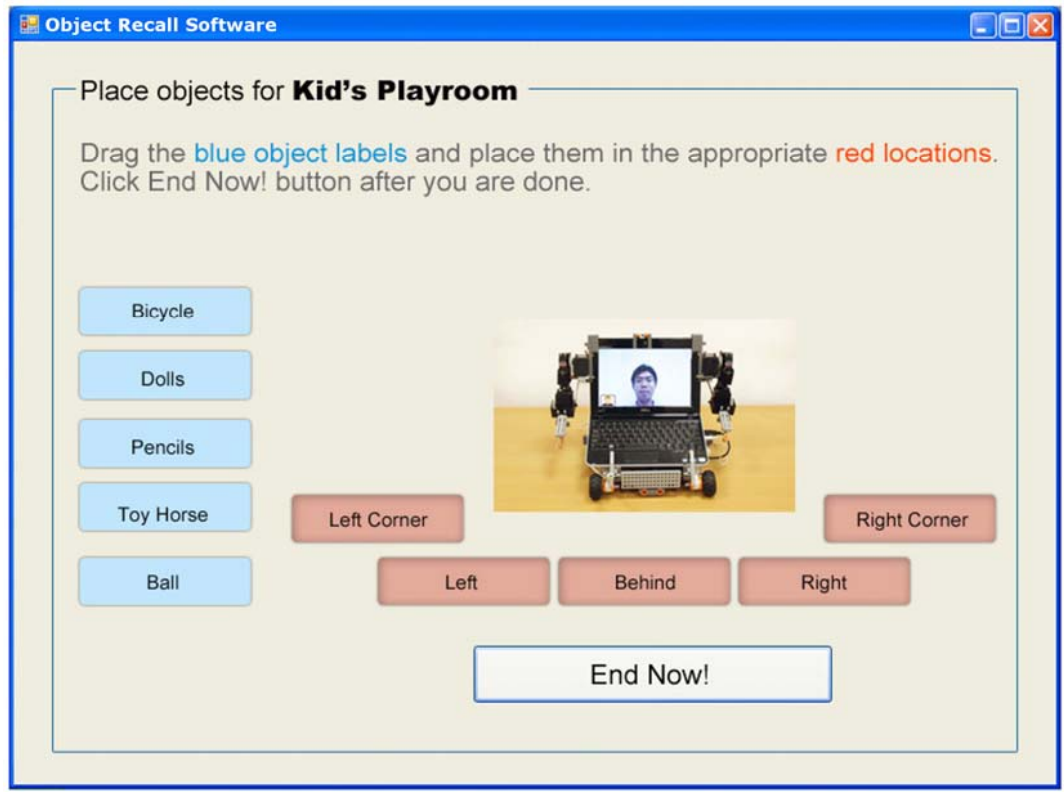

Fig. 3. A screenshot of the software that the participants used to drag-and-drop the objects to their corresponding locations.

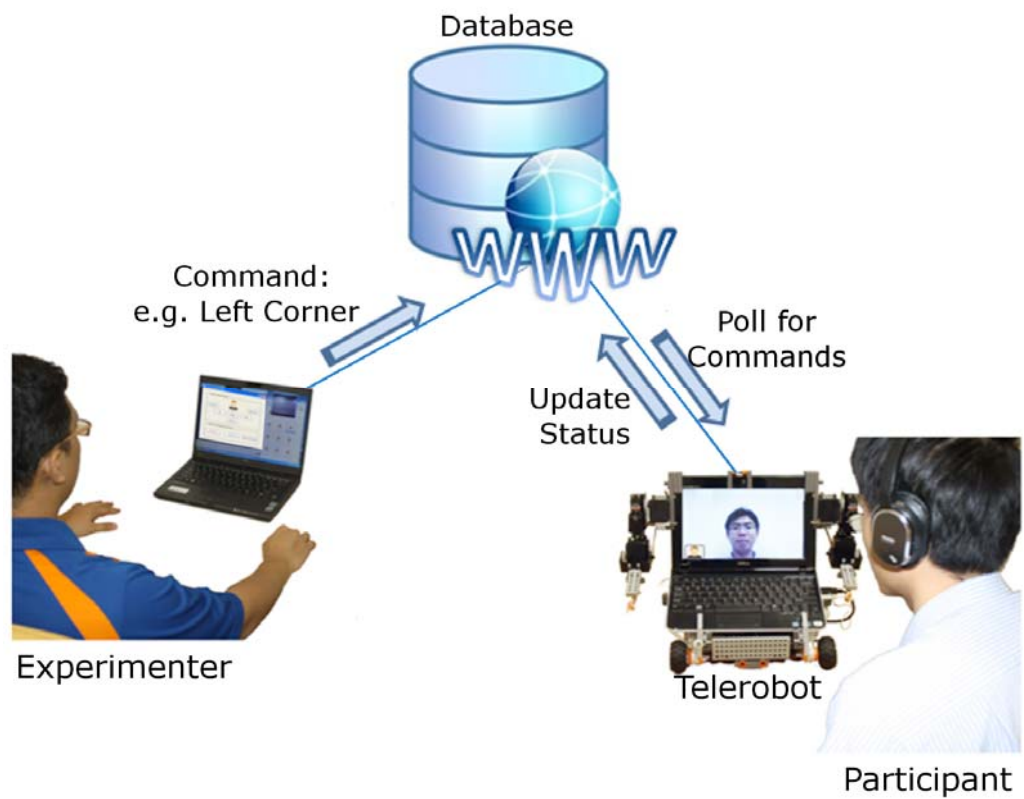

Fig. 4. The schematic diagram for controlling the telerobot's pointing arm.

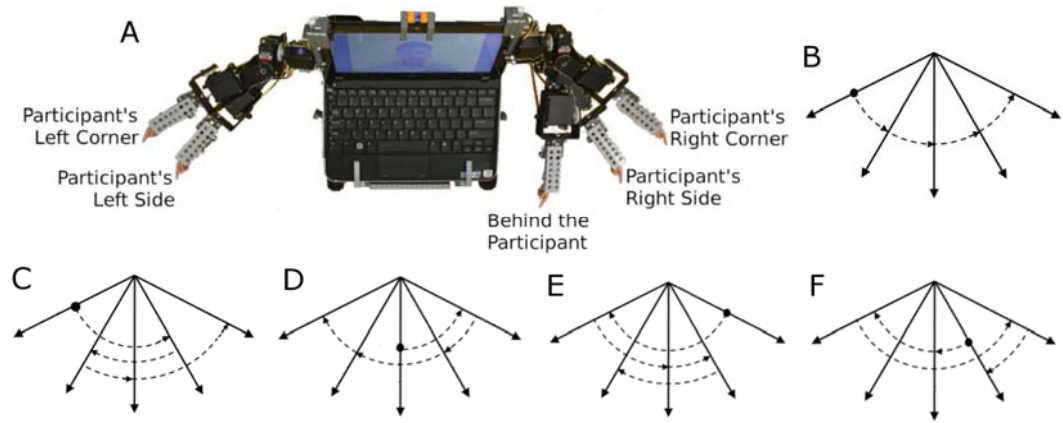

Fig. 5. The directions of the pointing gestures. (A) The telerobot with the pointing directions relative to the participant. The angular variations for (B) sequential order, 120°; and non-sequential order with pointing gestures for the  (C) kid's playroom, 210°; (D) kitchen, 180°; (E) living room, 270° and (F) bedroom, 270°.

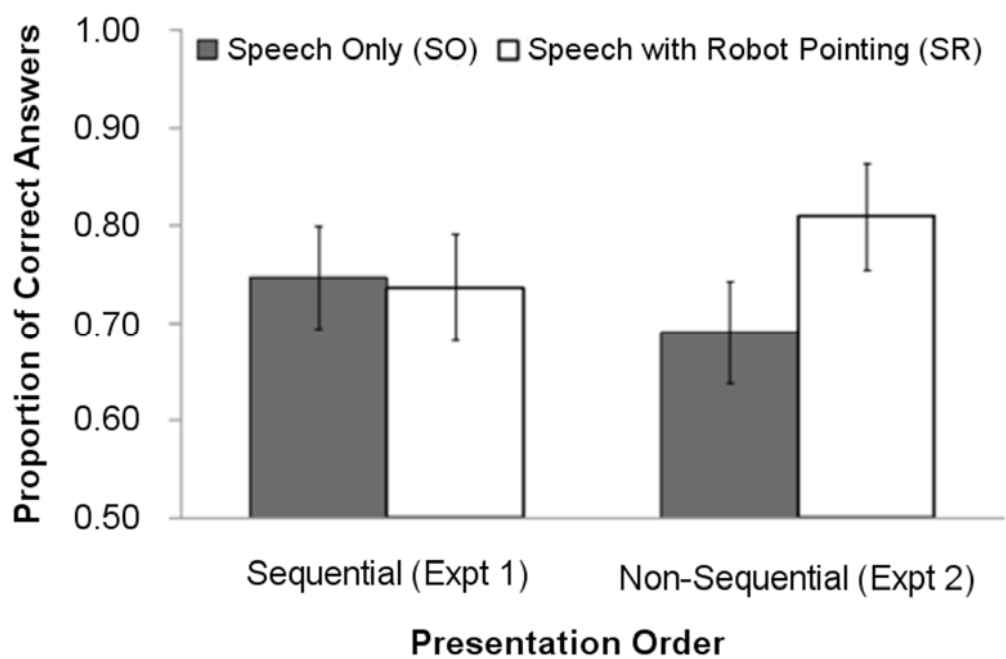

Fig. 6. Proportion of correct answers when fictitious objects were presented sequentially (N = 40) and non-sequentially (N = 42); error bars denote standard errors.

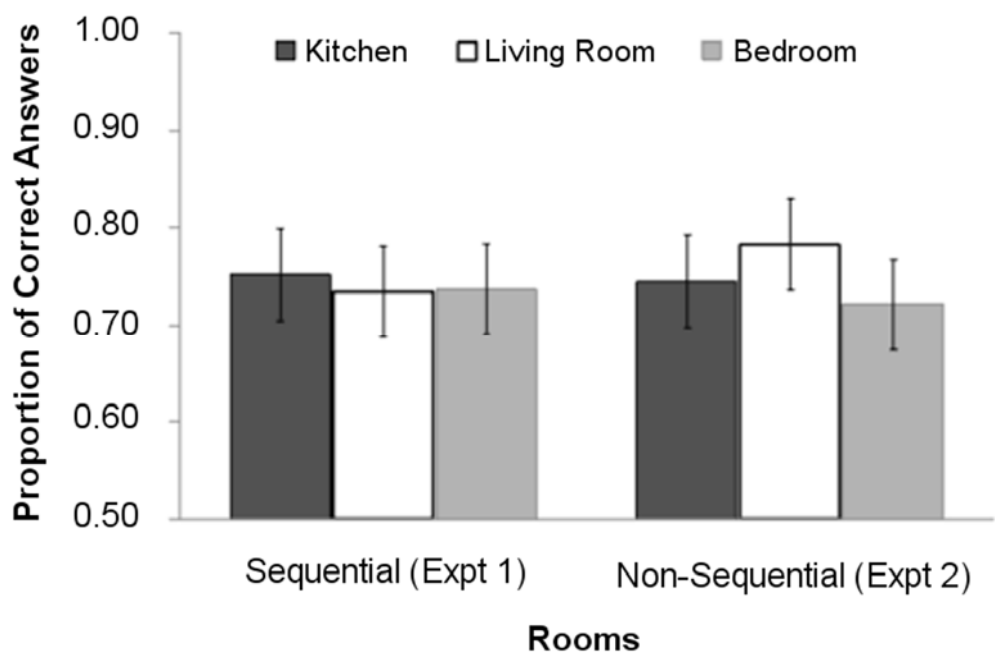

Fig. 7. Proportion of correctly recalled answers for the rooms. Each room corresponds to a controlled number of words.

# Bio

**John-John Cabibihan** was conferred with his PhD in biomedical robotics from the Scuola Superiore Sant'Anna, Pisa, Italy in 2007. He is an Assistant Professor at the Department of Electrical and Computer Engineering of the National University of Singapore. Concurrently, he serves as the Deputy Director of the Social Robotics Laboratory; Associate Editor of the International Journal of Social Robotics; and the Chair of the IEEE Systems, Man and Cybernetics Society (Singapore Chapter; terms: 2011 and 2012). He was the Program Co-Chair of the 2010 International Conference on Social Robotics, Singapore and Program Chair of the 2012 International Conference on Social Robotics, Chengdu, China. He is working on the core technologies towards lifelike touch and gestures for prosthetics and social robotics.

**Wing-Chee So** received her Ph.D. in the Department of Psychology at University of Chicago in 2007. Currently, she is working as an Assistant Professor at the Department of Educational Psychology in the Chinese University of Hong Kong. She is specialized in examining the role of gesture in communication and cognitive processes.

**Sujin Saj** graduated from the National University of Singapore in 2011 with a Bachelors degree in Engineering (Electrical) with a minor in Technology Entrepreneurship. He is currently working as a technology analyst in the financial industry focused primarily on banking technology infrastructure. He has a passion for futuristic technologies.

**Zhengchen Zhang** received the Bachelors degree in Engineering from the University of Science and Technology of China in 2004 and the Masters of Engineering degree from the Institute of Software, Chinese Academy of Sciences in 2008. He is currently working toward a Ph.D. degree in the Department of Electrical and Computer Engineering, Na- tional University of Singapore. His research interests include visual and audio communications in social robotics, text processing, and ma- chine learning.

# Appendix

|  |  | Presentation Order | |
|---|---|---|---|
|  |  | Sequential (Experiment 1) | Non-Sequential (Experiment 2) |
| **Presentation Mode** | **Speech Only (SO)** | Welcome to the kid's playroom. At your left corner is a *ball*. A *bicycle* is on your left side. The *dolls* are behind you. The *pencils* are on your right side. At your right corner is a *toy horse*. | Welcome to the kid's playroom. At your right corner is a *toy horse*. At your left corner is a *ball*. A *bicycle* is on your left side. The *pencils* are on your right side. The *dolls* are behind you. |
|  |  | Let me show you my kitchen. At your left corner is a *shopping bag*. The *refrigerator* is at your left. The *stove* is behind you. A *basket* is at your right. At your right corner are the *dishes*. (See **Movie S1**). | Let me show you my kitchen. The *refrigerator* is at your left. A *basket* is at your right. At your left corner is a *shopping bag*. The *stove* is behind you. At your right corner there are *dishes*. (See **Movie S3**). |
|  |  | Let me describe my living room to you. The *aquarium* is at your left corner. A *jar* is on your left. Behind you are the *DVDs*. A *couch* is on your right. The *stairs* are at your right corner. | Let me describe my living room to you. The *aquarium* is at your left corner. Behind you are the *DVDs*. The *stairs* are at your right corner. A *jar* is on your left. A *couch* is on your right. |
|  |  | Let's go to the bedroom. At your left corner is the *computer*. The *door* is at your left side. The *paintings* are behind you. The *bed* is at your right side. At your right corner is the *carpet*. | Let's go to the bedroom. The *bed* is at your right side. The *door* is at your left side. The *paintings* are behind you. At your right corner is the *carpet*. At your left corner is the *computer*. |
|  | **Speech with Robot Pointing (SR)** | Welcome to the kid's playroom. There are the *coloring books* (robot points to the participant's left corner). I see the *building blocks* over there (left side). That is a small *guitar* there (behind the participant). I can see a large *toy house* there (right side). You can find a *drawing board* over there (right corner). | Welcome to the kid's playroom. There are the *coloring books* (left corner). I can see a large *toy house* there (right side). I see the *building blocks* over there (behind). You can find a *drawing board* over there (right corner). |
|  |  | Let me now show you my kitchen. The *microwave oven* is over there (left corner). The *fruits* are just there (left side). You will find the *trashcan* over there (behind). There is the *barbecue stand* (right side). Over there I can see the *water dispenser* (right corner). (See **Movie S2**). | Let me now show you my kitchen. You will find the *trashcan* over there (behind). Over there I can see the *water dispenser* (right corner). There is the *barbecue stand* (right side). The *fruits* are just there (left side). The *microwave oven* is over there (left corner). (See **Movie S4**). |
|  |  | Now, let me describe my living room to you. My *piano* is over there (left corner). The *fire place* is just there (left side). The *ceiling fan* is there (behind). I can see the *sofa* over there (right side). I can see a *large window* there (right corner). | Now, let me describe my living room to you. I can see a large *window* there (right corner). My *piano* is over there (left corner). The *ceiling fan* is there (behind). I can see the *sofa* over there (right side). The *fire place* is just there (left side). |
|  |  | Let's go to the bedroom. Over there I can see several *candles* (left corner). I can see my new *curtain* on that side (left side). The *lamp shade* is just there (behind). The *cabinet* is on that side (right side). There are *clothes* over there (right corner). | Let's go to the bedroom. I can see my new *curtain* on that side (left side). The *lamp shade* is just there (behind). There are *clothes* over there (right side). Over there I can see several *candles* (left corner). The *cabinet* is on that side (right side). |